\documentclass[runningheads,envcountsect,envcountsame]{llncs}
\usepackage[T1]{fontenc}
\usepackage{fixltx2e}
\usepackage{textcomp}
\usepackage{epsf}
\usepackage[dvips]{epsfig}
\usepackage{amssymb,amsmath,amsfonts,authblk}
\usepackage{srcltx}
\usepackage{graphicx}
\usepackage{epsfig}
\usepackage{graphics}
\usepackage{gastex}
\usepackage[usenames]{color}
\title{
Canonical representatives of morphic permutations}
\author{Sergey V. Avgustinovich\inst{1}, Anna E. Frid\inst{2}, Svetlana Puzynina\thanks{Supported by the LABEX MILYON
(ANR-10-LABX-0070) of Universit\'{e} de Lyon, within the program
``Investissements d'Avenir'' (ANR-11-IDEX-0007) operated by the
French National Research Agency (ANR).}\inst{1,3}}
\institute{Sobolev Institute of Mathematics, Russia,
\email{avgust@math.nsc.ru} \and Aix-Marseille Universit\'{e},
France, \email{anna.e.frid@gmail.com} \and LIP, ENS de Lyon,
Universit\'e de Lyon, France, \email{s.puzynina@gmail.com}}
\begin{document}
\maketitle
\begin{abstract} An \emph{infinite permutation} can be defined as a linear ordering of the set of natural numbers.
In particular, an infinite permutation can be constructed  with an
aperiodic infinite word over $\{0,\ldots,q-1\}$ as the
 lexicographic order of the shifts of the word. In
this paper, we discuss the question if  an infinite 
permutation defined this way admits a {\em canonical}
representative, that is, can be defined by a sequence of numbers
from $[0, 1]$, such that the frequency of its elements in any
interval is equal to the length of that interval. We show
that 
a canonical representative exists if and only if the word is
uniquely ergodic, and that is why we use the term \emph{ergodic}
permutations. We also discuss ways to construct the canonical
representative of a permutation defined by a morphic word and
generalize the construction of Makarov, 2009, for the Thue-Morse
permutation to a wider class of infinite words.

\end{abstract}

\section{Introduction}
We continue the study of combinatorial properties of infinite permutations analogous to those of words.
In this approach, infinite permutations are interpreted as equivalence classes of real sequences with distinct
elements, such that only the order of elements is taken into account.
In other words, an infinite permutation is a linear order in $\mathbb N$. We consider it as an object close to an infinite word, but instead of symbols, we have transitive relations $<$ or $>$ between each pair of elements.

Infinite permutations in the considered sense were introduced in
\cite{ff}; see also a very similar approach coming from dynamics
\cite{bkp} and summarised in \cite{a}. Since then, they were
studied in two main directions: First, a series of results
compared properties of infinite permutations with those of
infinite words (\cite{ff,afks,f_fw} and others). Secondly,
different authors studied permutations directly constructed with
the use of general words \cite{elizalde,mak1}, as well as precise
examples: the Thue-Morse word \cite{mak_tm,wid1}, other morphic
words \cite{val,wid2} or Sturmian words \cite{mak_st}.

In the previous paper \cite{dlt}, we introduced the notion of an
\emph{ergodic} permutation, which means that a permutation can be
defined by a sequence of numbers from $[0,1]$ such that the
frequency of its elements in any interval is equal to the length
of the interval. We proved also that the minimal complexity (i.e.,
the number of subpermutations of length $n$) of an ergodic
permutation is $n$, and the permutations of minimal complexity are
Sturmian permutations in the sense of \cite{mak_st} (and close to
the sense of \cite{afks}). So, the situation for ergodic
permutations is similar to that for words. Note that for the
permutations in general, this is not the case: The complexity of
an aperiodic permutation can grow slower than any unbounded
growing function \cite{ff}.

In this paper, we focus on permutations generated by words. First
of all, we prove that such a permutation is ergodic if and only if
its generating word is uniquely ergodic, which explains the choice
of the term. Then we generalize the construction of Makarov
\cite{mak_tm} and give a general method to construct the canonical
representative sequence of any permutation generated by a fixed
point of a primitive monotone separable morphism. We also discuss
why this method cannot be directly extended further, and give some
examples.

\section{Basic definitions}
We consider finite and infinite words over a finite
alphabet $\Sigma_q = \{0, 1,q-1\}$. A \emph{factor}
 of an infinite word is any sequence of its
consecutive letters. The factor $u[i]\cdots u[j]$ of an infinite
word $u=u[0] u[1] \cdots u[n]
\cdots$, with $u[k] \in \Sigma$, is denoted by $u[i..j]$; \emph{prefixes} of a finite or an infinite word are as usual defined as starting factors.  

The
length of a finite word $s$ is denoted by $|s|$.
An infinite word $u=vww\cdots = vw^{\omega}$ for some non-empty
word $w$ is called \emph{ultimately $(|w|$-$)$periodic}; otherwise
it is called \emph{aperiodic}.

When considering words on $\Sigma_q$, we refer
to the \emph{order} on finite and infinite words meaning
lexicographic (partial) order: $0<1<\ldots<q-1$, and $u<v$ if $u[0..i]=v[0..i]$
and $u[i+1]<v[i+1]$ for some $i$. For words such that one of them
is the prefix of the other the order is not defined.


Now we recall the notion of the uniform frequency of letters and
factors in an infinite word. For finite words $v$ and $w$, we let
$|v|_w$ denote the number of occurrences of $w$ in $v$. The
infinite word $u$ has \emph{uniform frequencies} of factors if,
for every factor $w$ of $u$, the ratio $\frac{|u[i..i+n]|_w}{n+1}$ has a limit $\rho_w(u)$ when $n\to\infty$
uniformly in $k$. For more on uniform frequencies in words we
refer to \cite{CANT_freq}.


To define infinite permutations, we will use sequences of real
numbers. Analogously to a factor of a word, for a sequence
$(a[n])_{n=0}^{\infty}$ of real numbers, any of its finite
subsequences $a[i],a[i+1],\ldots,a[j]$ is called a \emph{factor}
and is denoted by $a[i..j]$. We define an equivalence relation
$\sim$ on real infinite sequences with pairwise distinct elements
as follows: $(a[n])_{n=0}^{\infty}\sim (b[n])_{n=0}^{\infty}$ if
and only if for all $i,j$ the conditions $a[i]<a[j]$ and
$b[i]<b[j]$ are equivalent. Since we consider only sequences of
pairwise distinct real numbers, the same condition can be defined
by substituting $(<)$ by $(>)$: $a[i]>a[j]$ if and only if
$b[i]>b[j]$. An \emph{infinite permutation} is then defined as an
equivalence class of real infinite sequences with pairwise
distinct elements. So, an infinite permutation is a linear
ordering of the set $\mathbb N_0=\{0,\ldots,n,\ldots\}$. We denote
it by $\alpha=(\alpha[n])_{n=0}^{\infty}$, where $\alpha[i]$ are
abstract elements equipped by an order: $\alpha[i] <\alpha[j]$ if
and only if $a[i]<a[j]$ or, which is the same, $b[i]<b[j]$ of
every \emph{representative} sequence $(a[n])$ or $(b[n])$ of
$\alpha$. So, one of the simplest ways to define an infinite
permutation is by a representative, which can be any sequence of
pairwise distinct real numbers.

\begin{example}\label{e1}
 Both sequences $(a[n])=(1,-1/2,1/4,\ldots)$ with $a[n]=(-1/2)^n$ and $(b[n])$ with $b[n]=1000+(-1/3)^n$ are representatives of the same permutation $\alpha=\alpha[0],\alpha[1],\ldots$ defined by
\[\alpha[2n]>\alpha[2n+2]>\alpha[2k+3]>\alpha[2k+1]\]
for all $n,k\geq 0$.
\end{example}

A \emph{factor} $\alpha[i..j]$ of an infinite permutation $\alpha$
is a finite sequence $(\alpha[i],\alpha[i+1],\ldots,\alpha[j])$ of
abstract elements equipped by the same order than in $\alpha$.
Note that a factor of an infinite permutation can be naturally
interpreted as a finite permutation: for example, if in a
representative $(a[n])$ we have a factor $(2.5, 2, 7,1.6)$, that
is, the 4th element is the smallest, followed by the 2nd, 1st and
3rd, then in the permutation, it will correspond to a factor
$\begin{pmatrix} 1 & 2 & 3 & 4 \\ 3 & 2 & 4 & 1 \end{pmatrix}$,
which we will denote simply as $(3241)$. Note that in general, we
index the elements of infinite objects (words, sequences or
permutations) starting with 0 and the elements of finite objects
starting with 1.

A factor of a sequence (permutation) should not be confused with its subsequence $a[n_0],a[n_1],\ldots$ (subpermutation $\alpha[n_0],\alpha[n_1],\ldots$) which is defined as indexed with a growing subsequence $(n_i)$ of indices.

Note, however, that in general, an infinite permutation cannot be defined as a permutation of $\mathbb N_0$. For instance, the permutation from Example \ref{e1} has all its elements between the first two ones.

\section{Ergodic permutations}

Let $(a[i])_{i=0}^{\infty}$ be a sequence of real numbers from the
interval $[0,1]$, representing an infinite permutation, $a$ and
$p$ also be real numbers from $[0,1]$. We say that the\emph{
probability} that an element of $(a[i])$ is less than $a$ exists and
is equal to $p$ if the ratio \[ \frac{ \# \{a[j+k]|0\leq k <n,
a[j+k]<a\}}{n}\] has a limit $p$ when $n \to \infty$ uniformly in
$j$.


In other words, if we substitute all the elements from $(a[i])$
which are smaller than $a$ by $1$, and those which are bigger by
$0$, the above condition means that the uniform frequency of the
letter $1$ exists and equals $p$. So, in fact the probability to
be smaller than $a$ is the uniform frequency of the elements which
are less than $a$.

We note that this is not exactly probability on the classical
sense, since we do not have a random sequence. But we are
interested in permutations where this ``probability'' behaves in
certain sense like the probability of a random sequence uniformly
distributed on $[0,1]$:

\begin{definition}{\rm A  sequence $(a[i])_{i=0}^{\infty}$ of real numbers is {\it canonical} if
\begin{itemize}
 \item
all the numbers are  pairwise distinct;
\item
for all $i$  we have $0\leq a[i] \leq 1$;
\item
 and for all $a$, the probability for any element $a[i]$ to be less than $a$ is well-defined and equal to $a$
 for all $a\in[0,1]$.
\end{itemize}
%
}
\end{definition}


\begin{remark}
 The set $\{a[i]|i \in \mathbb N\}$ for a canonical sequence $(a[i])$ is dense on $[0,1]$.
\end{remark}

\begin{remark}\label{r:int}
In a canonical sequence, the frequency of the elements which fall
into any interval $(t_1,t_2)\subseteq [0,1]$ exists and is equal
to $t_2-t_1$.
\end{remark}

\begin{remark}
Symmetrically to the condition ``the probability to be less than $a$ is $a$'' we can consider the equivalent condition ``the probability to be greater than $a$ is $1-a$''.
\end{remark}

\begin{definition}{\rm An infinite permutation $\alpha=(\alpha[i])_{i=1}^{\infty}$ is
called {\it ergodic} if it has a canonical
representative.}
\end{definition}

\begin{example}\label{ex:st}
 For any irrational $\sigma$ and for any $\rho$, consider the sequence of fractional parts $\{\rho+n\sigma\}$.
 It is uniformly distributed in $[0,1)$, so, the respective permutation is ergodic. In fact, such a permutation
 is a {\em Sturmian} permutation in the sense of \cite{mak_tm}; in \cite{afks}, the considered class of permutations
 is wider than that. It is easy to see that Sturmian permutations are directly related to Sturmian words
 \cite{Lo}.
\end{example}

\begin{proposition}
An ergodic permutation $\alpha$ has a unique canonical
representative.
\end{proposition}
\noindent {\sc Proof.} Given $\alpha$, for each $i$ we define
$$a[i]=\lim_{n\to \infty} \frac{\#\{\alpha[k]|0\leq k <n, \alpha[k]<\alpha[i]\}}{n}$$
and see that, first, this limit must exist since $\alpha$ is
ergodic, and secondly, $a[i]$ is the only possible value of an
element of a canonical representative of $\alpha$. \hfill $\Box$

\medskip
Note, however, that even if for some infinite permutation all the
limits above exist, it does not imply the existence of the
canonical representative. Indeed, there is another condition to
fulfill: for different $i$ the limits must be different.

\section{Ergodic permutations generated by words}

Consider an aperiodic infinite word $u=u[0]\cdots u[n] \cdots$
over $\Sigma_q$ and, as usual, define its $n$th shift $T^nu$ as
the word obtained from $u$ by erasing the first $n$ symbols: $T^n
u = u[n] u[{n+1}] \cdots$. We can also interpret a word $u$ as a
real number $0.u$ in the $q$-ary representation.

If the word $u$ is aperiodic, then in the sequence $(0.T^n
u)_{n=0}^\infty$ all the numbers are different and thus this
sequence is a representative of a permutation which we denote by
$\alpha_u$. Clearly, $\alpha_u[i]<\alpha_u[j]$ if and only if $T^i
u$ is lexicographically smaller than $T^j u$. A permutation which
can be constructed like this is called {\em valid}; the structure
of valid permutations has been studied in \cite{mak1} (for the
binary case) and \cite{elizalde} (in general).

Most of results of this paper were inspired by the following construction.

\begin{example}\label{e:tm}
The famous Thue-Morse word $0110100110010110\cdots$ is defined as the fixed point starting with 0 of the morphism $f_{tm}: 0 \mapsto 01, 1 \mapsto 10$. The respective Thue-Morse permutation defined by the representative ($0.01101001\cdots$, $0.11010011\cdots$, $0.10100110\cdots$, $0.01001100\cdots$,$\ldots$) can also be defined by the following sequence, denoted by $a_{tm}$:
\[\frac{1}{2},1,\frac{3}{4},\frac{1}{4},\frac{5}{8},\frac{1}{8},\frac{3}{8},\frac{7}{8},\cdots,\]
that is the fixed point of the  morphism $\varphi_{tm}: [0,1]
\mapsto [0,1]^2$:
\begin{equation*}
 \varphi_{tm}(x)=\begin{cases}
                                                          \frac{x}{2}+\frac{1}{4}, \frac{x}{2}+\frac{3}{4}, \mbox{~if~} 0 \leq x \leq \frac{1}{2},\\
                             \frac{x}{2}+\frac{1}{4}, \frac{x}{2}-\frac{1}{4}, \mbox{~if~} \frac{1}{2} < x \leq 1.
                                                         \end{cases}
\end{equation*}
It will be proved below that the latter sequence is canonical and
thus the Thue-Morse permutation is ergodic. This construction and
the equivalence of the two definitions was proved by Makarov in
2009 \cite{mak_st}; then the properties of the Thue-Morse
permutation were studied by Widmer \cite{wid1}.
\end{example}

When is a valid permutation ergodic? The answer is simple and explains the choice of the term ``ergodic''.

\begin{lemma}\label{l:erg}
A valid permutation $\alpha_u$ for a recurrent non-periodic word
$u$ is ergodic if and only if all the uniform frequencies of
factors in $u$ exist and are not equal to 0.
\end{lemma}

Before proving the lemma, we prove 
the following proposition about words: 

\begin{proposition}\label{c:maxmin} 
Let $u$ be a recurrent aperiodic word and $w$ and $v$ some of its
factors. Then in the orbit of $w$ there can be the
lexicographically maximal word from its closure starting with $w$,
or the lexicographically minimal word from its closure starting
with $v$, but not both at a time.
\end{proposition}
\noindent {\sc Proof.} Suppose the opposite: let $T^k(u)$ be the
maximal element of the orbit closure of $u$ starting with $w$, and
$T^l(u)$ be the minimal element of the orbit closure of $u$
starting with $v$. Consider the prefix $r$  of $u$ of length
$\max(k+|u|, l+|v|)$. Since $u$ is recurrent, this prefix appears
in it an infinite number of times, and since $u$ is not ultimately
periodic, there exists an extension $p$ of $r$ to the right which
is right special: $pa$ and $pb$ are factors of $u$ for some
symbols $a \neq b$. Suppose that the prefix of $u$ of the
respective length is $pa$, and $pb$ is a prefix of $T^n(u)$.

If $a<b$, then $u < T^n(u)$ and thus $T^k(u)< T^{k+n}(u)$, where
$T^{k+n}(u)$ starts with $w$. A contradiction with the maximality
of $T^k(u)$. If by contrary $a>b$, then $u > T^n(u)$ and thus
$T^l(u)> T^{l+n}(u)$, where $T^{l+n}(u)$ starts with $v$. A
contradiction with the minimality of $T^l(u)$. The proposition is
proved. \hfill $\Box$
 .

\medskip
\noindent {\sc Proof of Lemma \ref{l:erg}.}


Suppose first that the frequency $\mu(w)$ of each factor $w$ in
$u$ exists and is non-zero. We should prove that the corresponding
valid permutation is ergodic. For every $k$ we define
\[a[k]=\lim_{n \to \infty} \sum_{\substack{|v|=n,\\ v \leq w[k]\cdots w[{k+n-1}]}} \mu(v).\]
Clearly, such a limit exists and is in $[0,1]$, and by the
definition, the probability that another element of the sequence $(a[i])$ is less
than $a[k]$ is equal to $a[k]$.


It remains to prove that $a[k]\neq a[l]$ for $k \neq l$, that is,
that the sequence $(a[n])$ is indeed a representative of a
permutation.

Suppose the opposite: $a[k]=a[l]$ for $k \neq l$. Let $m\geq 0$ be
the first position such that $w[{k+m}]\neq w[{l+m}]$: say,
$w[{k+m}]<w[{l+m}]$. The only possibility for $a[l]$ and $a[k]$ to
be equal is that  $T^k(w)=w[k] w[{k+1}]\cdots$ is the maximal word
in the orbit closure of $w$ starting with $w[k]\cdots w[{k+m}]$,
and $T^l(w)=w[l] w[{l+1}]\cdots$ is the minimal word in the orbit
closure of $w$ starting with $w[l]\cdots w[{l+m}]$. Due to
Proposition \ref{c:maxmin}, this is a contradiction. So, the
values $a[k]$ are indeed all different, and thus the permutation
is well-defined. Together with the condition on the probabilities
we proved above, we get that the corresponding valid permutation
is ergodic.

\medskip

The proof of the converse is split into two parts. First we prove
that for a valid ergodic permutation the frequencies of factors in
the corresponding word must exist, then we prove that they are
non-zero.

So, first we suppose that the frequencies of (some) factors of $w$
do not exist. We are going to prove that the permutation is not
ergodic, that is, that the canonical representative sequence
$(a[n])$ is not well-defined.

Let us take the shortest and lexicographically minimal factor $w$
whose frequency does not exist and consider the subsequence
$(a[{n_i}])$ of the sequence $(a[n])$ corresponding to suffixes
starting with $w$. The upper limit  of $(a[{n_i}])$ should be
equal to the sum of frequencies of the words of length $|w|$ less
than or equal to $w$, but since the frequency of $w$ is the only
one of them that does not exist, this limit
also does not exist. So, the sequence $(a[n])$ is not well-defined and hence the corresponding valid permutation is not ergodic. 


\smallskip

The remaining case is that of zero frequencies: Suppose that $w$
is the shortest and lexicographically minimal factor whose
frequency is zero, and consider again the subsequence $(a[{n_i}])$
of the sequence $(a[n])$ corresponding to suffixes starting with
$w$. The subsequence $(a[{n_i}])$ is infinite since $u$ is
recurrent, but all its elements must be equal: Their value is the
sum of frequencies of words of length $|w|$ lexicographically less
than $w$. So, the sequence $(a[n])$ does not correctly define a permutation,
and hence in the case of zero frequencies the corresponding valid permutation is not ergodic. 
\hfill $\Box$

\bigskip \noindent
We have seen above in Example \ref{ex:st} how the canonical
representatives of permutations corresponding to Sturmian words
are built.

\begin{example}
Let us continue the Thue-Morse example started above and prove
that the representative $a_{tm}$ is canonical. We should prove
that the probability for any element $a[j]$ to be less than $a$ is
well-defined and equal to $a$.  Let us prove by induction on $k$
that the probability for an element to be in any binary rational
interval $(d/2^{k},(d+1)/2^k]$, where $0\leq d < 2^k$, is exactly
$1/2^k$. Indeed, by the construction, the intervals $(0,1/2]$ and
$(1/2,1]$ correspond to the zeros and ones in the original
Thue-Morse word whose frequencies are $1/2$. The morphic image of
any of these intervals is, consecutively, two intervals: $(0,1/2] \mapsto (1/4,2/4],(3/4,4/4]$, and 
$(1/2,1] \mapsto (2/4,3/4],(0,1/4]$. So, in both cases, the intervals are
 of the form
$(d/2^2,(d+1)/2^2]$, $d=0,\ldots,3$. Each of them is twice rarer
than its pre-image; the four intervals cover $(0,1]$ and do not
intersect, so, the probability for a point $a[i]$ to be in each of
them is $1/4$. But exactly the same argument works for any of
these four intervals: its image is two intervals which are twice
smaller and twice rarer than the pre-image interval. No other
points appear in that shorter interval since each mapping
corresponding to a position in the morphism is linear, and their
ranges do not intersect. So, the probability for a point to be in
an interval $(d/2^3,(d+1)/2^3]$ is $1/8$, and so on. By induction,
it is true for any binary rational interval and thus for all
interval subsets of $(0,1]$: the frequency of elements in this
interval is equal to its length. This proves that $a_{tm}$ is
indeed the canonical representative of the Thue-Morse permutation.
\end{example}

\begin{remark}
 {\rm This example shows that 
 the natural way of constructing the canonical representative of
 a valid permutation has little  in common with frequencies of factors in the underlying word. The frequencies of symbols look important, but, for example, the frequency of $00$ in the Thue-Morse word is $1/6$, whereas all the elements of the canonical representative are binary rationals.}
\end{remark}

\begin{remark}
 {\rm In Lemma \ref{l:erg}, we assumed that the word is recurrent. Indeed, if a word is not recurrent,
  the permutation can be ergodic. As an example, consider the word
\[01221211221121221\cdots,\]
that is, $0$ followed by the Thue-Morse word on the alphabet $\{1,2\}$. The respective permutation is still ergodic with the canonical representative $0,a_{tm}=0,1/2,1,3/4,1/4,\ldots$.

Note also that this property depends on the order of symbols. For example, the permutation associated with the word
\[20110100110010110\cdots=2 u_{tm}\]
is not ergodic since $a_tm[0]$ 
can be equal only to 1. On the other hand, it is well known that
the first shift of the Thue-Morse word is the lexicographically
largest element in its shift orbit closure. So, $a_tm[1]$ must
also be equal to 1. 

}
\end{remark}

\subsection{Morphisms on words and intervals}
In this subsection, we generalize the above construction 
for the Thue-Morse word  to a class of fixed points
of morphisms: for any word from that class, we construct a
morphism similar to the Thue-Morse interval morphism $\varphi_{tm}$ defined in Example \ref{e:tm}.

Let $\varphi: \{0,\ldots,q-1\}^*\mapsto \{0,\ldots,q-1\}^*$ be a
morphism and $u=\varphi(u)$ be its aperiodic infinite fixed point
starting with a letter $a$ if it exists. In what follows we give a
construction of the canonical representative $a_u$ of the
permutation $\alpha_u$ provided that the morphism $\varphi$ is
{\it primitive, monotone} and {\it separable}. We will now define
what these properties mean.

Recall that the matrix $A$ of a morphism $\varphi$ is a $q\times
q$-matrix whose element $a_{ij}$ is equal to the number of
occurrences of $i$ in $\varphi(j)$. A matrix $A$ and a morphism
$\varphi$ are called {\it primitive} if in some power $A^n$ of $A$
all the entries are positive, i.e.,  
for every $b \in \{0,\ldots,q-1\}$ all the symbols of
$\{0,\ldots,q-1\}$ appear in $\varphi^n(b)$ for some
$n$. A classical Perron-Frobenius theorem says that a primitive
matrix has a dominant positive {\it Perron-Frobenius eigenvalue}
$\theta$ such that $\theta>|\lambda|$ for any other eigenvalue
$\lambda$ of $A$. It is also well-known 
that a fixed
point of a primitive morphism is uniquely ergodic, and that the
vector $\mu =(\mu(0),\ldots,\mu(q-1))^t$ of frequencies of symbols
is the normalized Perron-Frobenius eigenvector of $A$:
\[A\mu=\theta \mu.\]

We say that a morphism $\varphi$ is {\it monotone on an infinite
word} $u$ if for any $n,m>0$ we have $T^n(u)<T^m(u)$ if and only
if $\varphi(T^n(u))<\varphi(T^m(u))$; here $<$ denotes the
lexicographic order. A morphism is called {\it monotone} if it is
monotone on all infinite words, or, equivalently, if for any
infinite words $u$ and $v$ we have $u<v$ if and only if
$\varphi(u)<\varphi(v)$.
\begin{example}
 {\rm The Thue-Morse morphism $\varphi_{tm}$ is monotone since $01=f_{tm}(0)<f_{tm}(1)=10$.  }
\end{example}
\begin{example}\label{ex:fib2}
 {\rm The Fibonacci morphism $\varphi_f: 0 \to 01, 1 \to 0$ is not monotone since
 $01=\varphi_f(0)>\varphi_f(10)=001$, whereas $0<10$. At the same time, $\varphi_f^2: 0 \to 010, 1 \to 01$
 is monotone since for all $x,y\in\{0,1\}$ we have $\varphi^2_f(0x)=0100x'<0101y'=\varphi_f(1y)$, where $x',y'\in\{0,1\}^*$. So, to use our construction to the Fibonacci word $u_f=01001010\cdots$ which is the fixed point of $\varphi_f$, we should consider $u_f$ as the fixed point of $\varphi_f^2$.}
\end{example}
\begin{example}\label{ex:nonmon}
 {\rm As an example of a morphism which does not become monotone even when we consider its powers, consider $g: 0 \to 02, 1 \to 01, 2 \to 21$. It can be easily seen that $g^n(0)>g^n(1)$ for all $n \geq 1$.
 }
\end{example}
The last condition we require from our morphism is to be {\it
separable}. To define this property, consider the fixed point $u$
as the infinite catenation of morphic images of its letters and
say that the {\it type} $\tau(n)$ of a position $n$ is the pair
$(a,p)$ such that $u[n]=\varphi(a)[p]$ in this ``correct''
decomposition into images of letters. So, there are
$\sum_{a=0}^{q-1} |\varphi(a)|$ different types of positions in
$u$. 
Also note that we index the elements of $u$ starting with 0 and
the elements of finite words $\varphi(a)$ starting from 1, so
that, for example,
$\tau(0)=(u[0],1)$. 

We say that a fixed point $u$ of a morphism $\varphi$ is {\it
separable} if for every $n,m$ such that $\tau(n)\neq \tau(m)$ the
relation between $T^n(u)$ and $T^m(u)$ is uniquely defined by the
pair $\tau(n), \tau(m)$. For a separable morphism $\varphi$ we
write $\tau(n)\preceq \tau(m)$ if and only if $T^n(u) \leq
T^m(u)$.

\begin{example}\label{tm_order}
 {\rm The Thue-Morse word is separable since for $\tau(n)=(0,1)$ and $\tau(m)=(1,2)$
 we always have $T^n(u_{tm})>T^m(u_{tm})$, i.e., 
 all zeros which are first symbols of $f_{tm}(0)=01$ give greater words than zeros which
 are second symbols of $f_{tm}(1)=10$.
Symmetrically, all ones which are first symbols of $f_{tm}(1)=10$
give smaller words than ones which are second symbols of
$f_{tm}(0)=01$, that is, for $\tau(n)=(1,1)$ and $\tau(m)=(0,2)$
we always have $T^n(u_{tm})<T^m(u_{tm})$.}
\end{example}
\begin{example}\label{ex:nonsep}
 {\rm
The fixed point
\[u=001001011001001011001011011\cdots\]
 of the morphism $0 \to 001, 1 \to 011$ is inseparable. Indeed, compare the following shifts: $T^2(u)=1001011001\cdots$,
$T^5(u)=1011\cdots$ and $T^{17}(u)=1001011011\cdots$. We see that $T^2(u)<T^{17}(u)<T^5(u)$. 
At the same time, $\tau(2)=\tau(5)=(0,3)$, and $\tau(17)=(1,3)$.
}
\end{example}
Note that the class of primitive monotone separable morphisms
includes in particular all morphisms considered by Valyuzhenich
\cite{val} who gave a formula for the permutation complexity of
respective fixed points.

\smallskip

Similarly to morphisms on words, we define \emph{a morphism on
sequences of numbers} from an interval $[a,b]$ as a mapping
$\varphi: [a,b]^*\mapsto [a,b]^*$. A fixed point of the morphism
$\varphi$ is defined as an infinite sequence $a[0], a[1], \dots $
of numbers from $[a,b]$, such that $\varphi (a[0], a[1], \dots) =
a[0], a[1], \dots $. Clearly, if a morphism $\varphi$ has a fixed
point, then there exists a number $c\in[a,b]$ such that $\varphi
(c) = c, c[1], \dots, c[k]$ for some $k\geq 1$ and $c[i] \in
[a,b]$ for $i=1, \dots k$. Clearly, a fixed point of a morphism on
sequences of numbers defines an infinite permutation (more
precisely, its representative) if and only if all the elements of
the sequence are distinct. The example of morphism defining an
infinite permutation is given by the Thue-Morse permutation
described in Example \ref{e:tm}.

\smallskip

The rest of the section is organized as follows: First we provide
the construction of a morphic ergodic permutation, then we give
some examples, and finally we prove the correctness of the
construction.

\smallskip

\noindent \textbf{The constuction of ergodic permutation
corresponding to a separable fixed point of a monotone primitive
morphism.}

Now let us consider a separable fixed point $u$ of a monotone
primitive morphism $\varphi$ over the alphabet $\{0,\ldots,q-1\}$,
and construct the canonical representative $a_u$ of the
premutation $\alpha_u$ generated by it. To do it, we first look if
$u$ contains lexicographically minimal or maximal elements of the
orbit with a given prefix. Note that due to Proposition
\ref{c:maxmin}, it cannot contain both of them. So, if $u$ does
not contain lexicographically maximal elements, we consider all
the intervals to be half-open $[\cdot )$; in the opposite case, we
can consider them to be half-open $(\cdot ]$, like in the
Thue-Morse case. Without loss of generality, in what follows we
write the intervals $[\cdot )$, but the case of $(\cdot ]$ is
symmetric.

So, let $\mu=(\mu_0,\ldots,\mu_{q-1})$ be the vector of
frequencies of symbols in $u$. Take the intervals $I_0=[0,\mu_0)$,
$I_1=[\mu_0,\mu_0+\mu_1)$, $\ldots$, $I_{q-1}=[1-\mu_{q-1},1)$.
An element $e$ of $a_u$ is in $I_b$ if for another element of $a_u$ the probability  to be less than $e$ is greater than the sum of frequences of letters less than $b$, and the probability to be greater than $e$ is greater than the sum of frequences of letters greater than $b$. In other words, $e$ is in $I_b$ if and only if the
respective symbol of $u$ is $b$.


Now let us take all the $k=\sum_{a=0}^{q-1} |\varphi(a)|$ types of
positions in $u$ and denote them according to the order $\preceq$:
\[\tau_1\prec \tau_2 \prec \cdots \prec \tau_k\],
with $\tau_i=(a_i,p_i)$.


For each $\tau_i$ the frequency $l_i=\mu_{a_i}/\theta$, where $\theta$ is the Perron-Frobenius eigenvalue of $\varphi$, is the frequency of symbols of type $\tau_i$ in $u$. Indeed, the $\varphi$-images of $a_i$ are $\Theta$ times rarer in $u$ than $a_i$, and $\tau_i$ corresponds just to a position in such an image. Denote
\[J_1=[0,l_1), J_2=[l_1,l_1+l_2),\ldots, J_k=[1-l_k,1);\]
so that in general, $J_i=[\sum_{m=1}^{i-1}l_m,\sum_{m=1}^{i}l_m)$.
We will also denote $J_i=J{a_i,p_i}$.

The interval $J_i$ is the range of elements of $a_u$ corresponding
to the symbols of type $\tau_i$ in $u$. Note that all symbols of
the same type are equal, and on the other hand, each symbol is of some type. For example, we have a collection of possible positions of 0 in images of letters, that is, a collection of types corresponding to 0, and all these types are less than any other type corresponding to any other symbol. So, the union of elements $J_i$ corresponding to 0 is exactly $I_0$, and the same argument can be repeated for any greater symbol. In particular, each $J_i$
is a subinterval of some $I_{a}$.

Now we define the morphism $\psi: [0,1]^* \mapsto [0,1]^*$ as
follows: For $x \in I_a$ we have
\[\psi(x) = \psi_{a,1}(x),\ldots \psi_{a,|\varphi(a)|}.\]
Here $\psi_{a,p}$ is a linear mapping $\psi_{a,p}:I_a \mapsto
J_{a,p}$: If $I_a=[x_1,x_2)$ and $J_{a,p}=[y_1,y_2)$, then
\begin{equation}\label{e:psi}
\psi_{a,p}(x)=\frac{y_2-y_1}{x_2-x_1}(x-x_1)+y_1.
\end{equation}

Now we can define the starting point, that is, the value of $a_1$.
Suppose that the first symbol of $u$ is $b$; then $\varphi(b)$
starts with $b$, which means that $J_{b,1}\subset I_b$, and the
mapping $\psi_{b,1}$ has a fixed point {$x$: $\psi_{b,1}(x)=x$. We
take $a_1$ to be this fixed point: $a_1=x$. Note that if $a_1$ is
the upper end of $J_{b,1}$, then we should take all the intervals
to be $(\cdot ]$; if it is the lower end, the intervals are
$[\cdot )$; if it is in the middle of the interval, the ends are
never attained. The situation when $a_1$ is an end of $J_{b,1}$ corresponds to the situation when there are the least or the greatest infinite words starting from some prefix in the orbit of $u$; as we have seen in Proposition \ref{c:maxmin}, only one of these situations can appear at a time. In particular, in this situation, $u$ is the least (or greatest) element of its orbit starting with $b$. 

This construction may look bulky, but in fact, it is just a
natural generalization of that for the Thue-Morse word. Indeed, in
the Thue-Morse word, $\mu_0=\mu_1=1/2$, $\theta=2$, and the order
of types is given in Example \ref{tm_order}. So, $I_0=[0,1/2]$,
$I_1=[1/2,1]$, $J_{0,1}=[1/4,1/2]$, $J_{0,2}=[3/4,1]$,
$J_{1,1}=[1/2,3/4]$, $J_{1,2}=[0,1/4]$. Here the intervals are
written as closed since at this stage we do not yet know whether
we must take them $[\cdot )$ or $( \cdot ]$. However, it becomes
clear as soon as we consider the mapping $\psi_{0,1}$ which is the
linear order-preserving mapping $I_0 \mapsto J_{0,1}$. Its fixed
point is 1/2, that is, the upper end of both intervals. Thus, the
intervals must be chosen as $(\cdot ]$. The mappings $\psi_{a,p}$
are explicitly written down in Example \ref{e:tm}.

\smallskip

To give another example, consider the square of the Fibonacci morphism mentioned in Example \ref{ex:fib2}.
\begin{example}\label{ex:fib3}
 {\rm
Consider the Fibonacci word as the fixed point of the square of the Fibonacci morphism: $\varphi^2_{f}: 0 \to 010, 1 \to 01$. This morphism is clearly primitive; also, it is monotone as we have seen in Example \ref{ex:fib2}, and separable: we can check that $(0,3)\preceq (0,1) \preceq (1,1) \preceq (0,2) \preceq (1,2)$. In particular, this means that zeros which are first symbols of $\varphi^2_f$ are in the middle among other zeros. So, in what follows we can consider open intervals since their ends are never attained.

The Perron-Frobenius eigenvalue is $\theta=(3+\sqrt{5})/2$, the frequencies of symbols are $\mu_0=(\sqrt{5}-1)/2$ and $\mu_1=(3-\sqrt{5})/2$. So, we have
\[I_0=\left(0,\frac{\sqrt{5}-1}{2}\right ),I_1=\left (\frac{\sqrt{5}-1}{2},1 \right),\]
and divide their lengths by $\theta$ to get the lengths of intervals corresponding to symbols from their images:
\[|J_{0,1}|=|J_{0,2}|=|J_{0,3}|=\frac{\mu_0}{\theta}=\sqrt{5}-2, \; |J_{1,1}|=|J_{1,2}|=\frac{\mu_1}{\theta}=\frac{7-3\sqrt{5}}{2}.\]
The order of intervals is shown at Fig.~\ref{f2}.

\begin{figure}
\centering \includegraphics[width=0.6\textwidth]{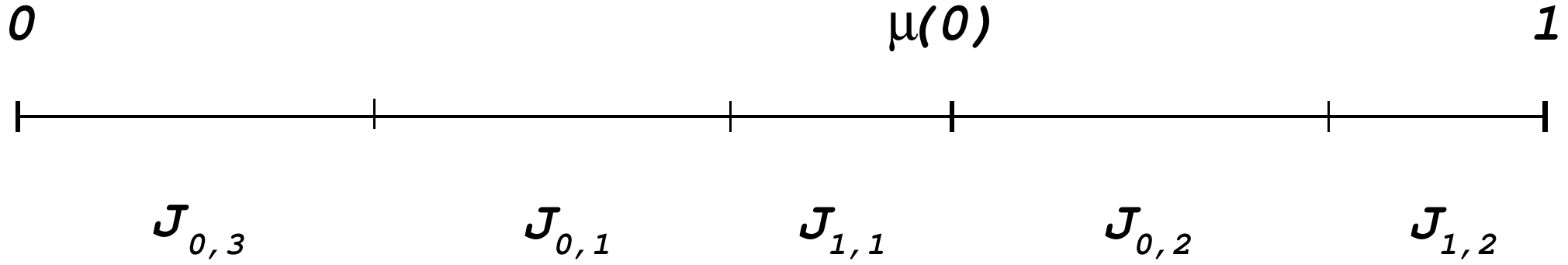}
\caption{Intervals for the Fibonacci permutation morphism}\label{f2}
\end{figure}

Now the morphism $\psi$ can be completely defined:
\[\psi(x)=\begin{cases}
           \psi_{0,1}(x),\psi_{0,2}(x),\psi_{0,3}(x) \mbox{~for~} x \in I_0,\\
           \psi_{1,1}(x), \psi_{1,2}(x) \mbox{~for~} x \in I_1.
          \end{cases}
\]
Here the mappings $\psi_{a,p}: I_a \mapsto J_{a,p}$ are defined according to \eqref{e:psi}. In particular, $\psi_{0,1}: (0,(\sqrt{5}-1)/2) \mapsto (\sqrt{5}-2,2(\sqrt{5}-2))$ has the fixed point $x=\psi_{0,1}(x)=(3-\sqrt{5})/2$. This is the starting point $a_1$ of the fixed point $a$ of $\psi$.
}
\end{example}

We remark that we could prove directly that the sequence $a$
constructed above is exactly the canonical representative of the
permutation associated with the Fibonacci word, using the fact
that Fibonacci word belongs to the family of Sturmian words.
However, we do not provide the proof for this example, since we
now give a more general proof of the correctness of the general
construction: the fixed point of the morphism $\psi$ described
above is indeed the canonical representative of our permutation.

\medskip

\noindent \textbf{Proof of correctness of the construction of the
morphism $\psi$.}

\smallskip

First we show that the fixed point of $\psi$ is a representative
of our permutation. Indeed, if $T^n(u)<T^m(u)$, and $n$ and $m$
are of different types, then, since the morphism is separable and
by the construction, $a[n]$ and $a[m]$ are in different intervals
$J_{a,p}$, and $a[n]<a[m]$. Now suppose that $n$ and $m$ are of
the same type $(a,p)$, that is, the $n$th ($m$th) symbol of $u$ is
the symbol number $p$ of the image $\varphi(a)$, where $a$ is the
symbol number $n'$ ($m'$) of $u$, i.e., $u[n']=a$,
$u[n]=\varphi(a)[p]$, and applying the morphism $\varphi$ to $u$
sends $u[n']$ to $u[n-p+1 .. n-p+|\varphi(a)|]$. Then, since the
morphism is monotone, $T^n(u)<T^m(u)$ if and only of
$T^{n'}(u)<T^{m'}(u)$. Exactly the same condition is true for the
relation $a_[n]<a_{m}$ if and only if $a_{n'}<a_{m'}$, since the
mapping $\psi_{a,p}$ preserves the order. Now we can apply the
same arguments to $m'$ and $n'$ instead of $m$ and $n$, and so on.
So, by the induction on the maximal power of $\varphi$ involved,
we also get that $T^n(u)<T^m(u)$ if and only if $a_[n]<a_[m]$. So,
the sequence $a$ is indeed a representative of the permutation
generated by $u$.

It remains to prove that this representative is canonical. As
above for the Thue-Morse word, it is done inductively on the
intervals 
\[\psi_{b_k,p_k}(\psi_{b_{k-1},p_{k-1}}( \ldots
\psi_{b_1,p_1}(I_{b_1})\ldots )).\] 
We prove that the probability
for an element of $a$ to be in this interval is equal to its
length. For the intervals $I_b$, it is true by the construction as
well as for their images. To make an induction step, we observe
that the image of an interval under each $\psi_{b,p}$ is $\theta$
times smaller than the initial interval and corresponds to the
situation which is $\theta$ times rarer. So, we have a partition
of $(0,1)$ to arbitrary small intervals for which the length is
equal to the frequency of occurrences. This is sufficient to make
sure that in fact, this is true for all intervals. \hfill $\Box$



\begin{remark}
 {\rm In Example \ref{ex:fib3}, we constructed a morphism for the Fibonacci permutation. 
However, it is not unique, and even not unique among piecewise linear morphisms. For example, 
the canonical representative $b$ of each Sturmian permutation $\beta(\sigma,\rho)$ defined by 
$\beta_n=\{\sigma n + \rho\}$ for $n \geq 0$ is the fixed point of the following morphism 
$[0,1]^* \mapsto [0,1]^*$:
$x \to \{2x-\rho\}, \{2x-\rho+\sigma\}.$
Indeed, this is exactly a morphism which sends $\{\sigma n + \rho\}$ to $\{\sigma (2n) + \rho\}$, $\{\sigma(2 n+1) + \rho\}$. It is clearly piecewise linear as well as the function $\{\cdot \}$. Also, the same idea can be generalized to a $k$-uniform morphism for any $k \geq 2$.
}
\end{remark}

\begin{remark}
We remark that the considerations used in the proof of the
correctness of the construction are closely related to so-called
Dumont-Thomas numeration systems \cite{DT}.

\end{remark}

\end{document}